\documentclass{PoS}

\def\z{\phantom{0}}
\def\nh{$N_{\mathrm H}$}

\title{AGN population in the deepest hard X-ray extragalactic survey}

\ShortTitle{AGN population in the deepest hard X-ray extragalactic survey}

\author{\speaker{St\'ephane Paltani}, Thierry J.-L. Courvoisier\\
        ISDC Data Centre for Astrophysics, Geneva Observatory, University of Geneva, Switzerland\\
        E-mail: \email{Stephane.Paltani@unige.ch}, \email{Thierry.Courvoisier@unige.ch}}

\author{Tom Dwelly, Ian M. McHardy\\
        School of Physics and Astronomy, University of Southampton, Southampton UK\\
        E-mail: \email{td@phys.soton.ac.uk}, \email{imh@astro.soton.ac.uk}}

\author{Roland Walter\\
        ISDC Data Centre for Astrophysics, Geneva Observatory, University of Geneva, Switzerland\\
        E-mail: \email{Roland.Walter@unige.ch}}

\abstract{We present the results of the analysis of the AGN population in the deepest extragalactic hard X-ray survey. The survey is based on INTEGRAL observation of the 3C 273/Coma cluster region, and covers 2500\,deg$^2$ with a 20--60\,keV flux limit 1.5 times lower than other surveys at similar energies, resolving about 2.5\% of the cosmic hard X-ray background. Using this survey, we can constrain in an unbiased way the distribution of hydrogen column absorption up to \nh$=10^{25}$\,cm$^{-2}$. We put an upper limit of 24\% to the fraction of Compton-thick objects. Compared to models of the AGN population selected in the 2--10\,keV band, the Log\,$N$--Log\,$S$ diagram is generally in good agreement, but the \nh\ distribution is significantly different, with significantly less unabsorbed sources (\nh$<10^{22}$\,cm$^{-2}$) at a given flux limit compared to the models. We also study the local hard X-ray luminosity function (LF), which is compatible with what is found in other recent hard X-ray surveys. The extrapolation of the 2--10\,keV LF is lower than the hard X-ray LF. The discrepancy is resolved if AGN spectra typically present reflection humps with reflection fraction $R\sim 1$. Finally, we use the population properties of this survey to show that a future ultra-deep INTEGRAL extragalactic survey can result in a quite large AGN sample with enough objects at redshifts larger than $z=0.05$ so that we can detect evolution in the hard X-ray LF.}

\FullConference{7th INTEGRAL Workshop\\
		 September 8-11 2008\\
		 Copenhagen, Denmark}

\begin{document}

\section{Introduction}
The observation of the central parts of local galaxies have shown that a tight connection exists between supermassive black holes and their host galaxies \cite{MagoEtal-1998-DemMas,FerrMerr-2000-FunRel,GebhEtal-2000-BlaHol}. Numerical simulations of galaxy mergers \cite{KaufHaeh-2000-UniMod,HopkEtal-2005-PhyMod} even suggest that the actively accreting 
supermassive black holes -- what we call active galactic nuclei (AGN) -- may play a fundamental role in the evolution of the galaxies. AGN surveys have therefore become an important part of observational cosmology.

The X-ray domain is very efficient at targeting AGN, in large part because the vast majority of high-latitude X-ray sources are AGN. However, it is expected that X-ray surveys may provide only a partial view of the AGN population.  The HEAO-1 satellite discovered an apparently diffuse X-ray emission at high galactic latitude \cite{MarsEtal-1980-DifXra} whose spectrum peaks around 30\,keV. While AGN could be the most important contributor to the so-called cosmic X-ray background, the spectrum could not be explained by the sum of AGN spectra, unless there exists a population of highly absorbed AGN, with hydrogen column densities \nh\ larger than $10^{22}$\,cm$^{-2}$, and even Compton-thick objects with \nh$\gtrsim 10^{24}$\,cm$^{-2}$  \cite{SettWolt-1989-ActGal,MadaEtal-1994-UniSey,MattFabi-1994-SpeCon,ComaEtal-1995-ConAGN}. These Compton-thick objects emit very little radiation below 10\,keV and thus require deep X-ray observations. In a recent detailed modeling of the AGN population based on the known AGN population up to $z\sim 3$ and its extrapolation to higher redshifts, \cite{GillEtal-2007-SynCos} found that the population of Compton-thick AGN should be as large as that of moderately absorbed AGN.

Absorption is much less efficient in the hard X-ray domain ($\gtrsim 20$\,keV) than in lower X-ray bands. Surveys in the hard X-ray have therefore the potential of detecting bright AGN with minimal bias in the \nh\ distribution below $\sim 10^{25}$\,cm$^{-2}$. INTEGRAL and SWIFT are two satellites with such survey capabilities which allowed to study the AGN population over the full sky \cite{SazoEtal-2007-HarXra,TuelEtal-2008-SwiBat}. However, the sensitivity limit of these surveys remains very low compared to those of surveys conducted by XMM-Newton or Chandra.

We present here the analysis of new INTEGRAL data covering a $\sim 2500$\,deg$^2$ region of the sky centered around 3C~273 and the Coma cluster. Thanks to these data, this survey is the deepest high-latitude survey in the hard X-rays. Most results have been presented in detail in \cite{PaltEtal-2008-DeeInt}. We report on the population properties of the detected AGN, and in particular their absorption properties compared to what is known from the AGN population observed in 2-10\,keV X-ray surveys. We show here what can be derived from the comparison of the LFs in the 2--10\,keV and 20--60\,keV ranges. Finally, we explore the usefulness of performing a much deeper INTEGRAL high-latitude survey in the form of a 5x5 dithering pattern with up to 20\,Ms total exposure time.

\section{The deepest INTEGRAL/IBIS extragalactic field}
\begin{figure}
\begin{center}
 \includegraphics[width=.7\textwidth,angle=90]{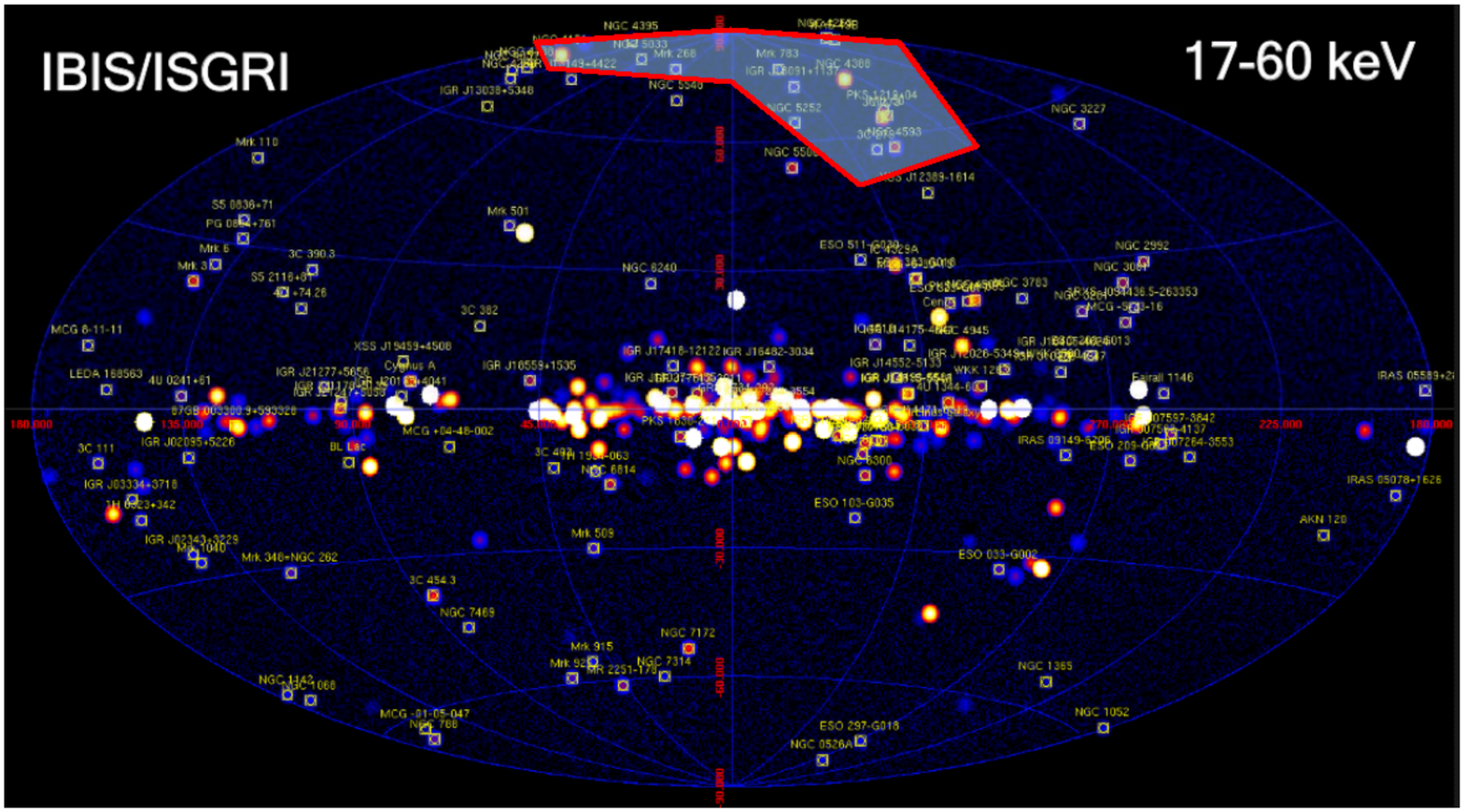}
 \caption{INTEGRAL/IBIS all-sky survey \cite{SazoEtal-2007-HarXra} showing the region of the deepest high-latitude INTEGRAL exposure analyzed in \cite{PaltEtal-2008-DeeInt} (blue area with red contours).}
 \label{survey}
\end{center}
\end{figure}
\begin{figure}
\begin{center}
 \includegraphics[width=.48\textwidth]{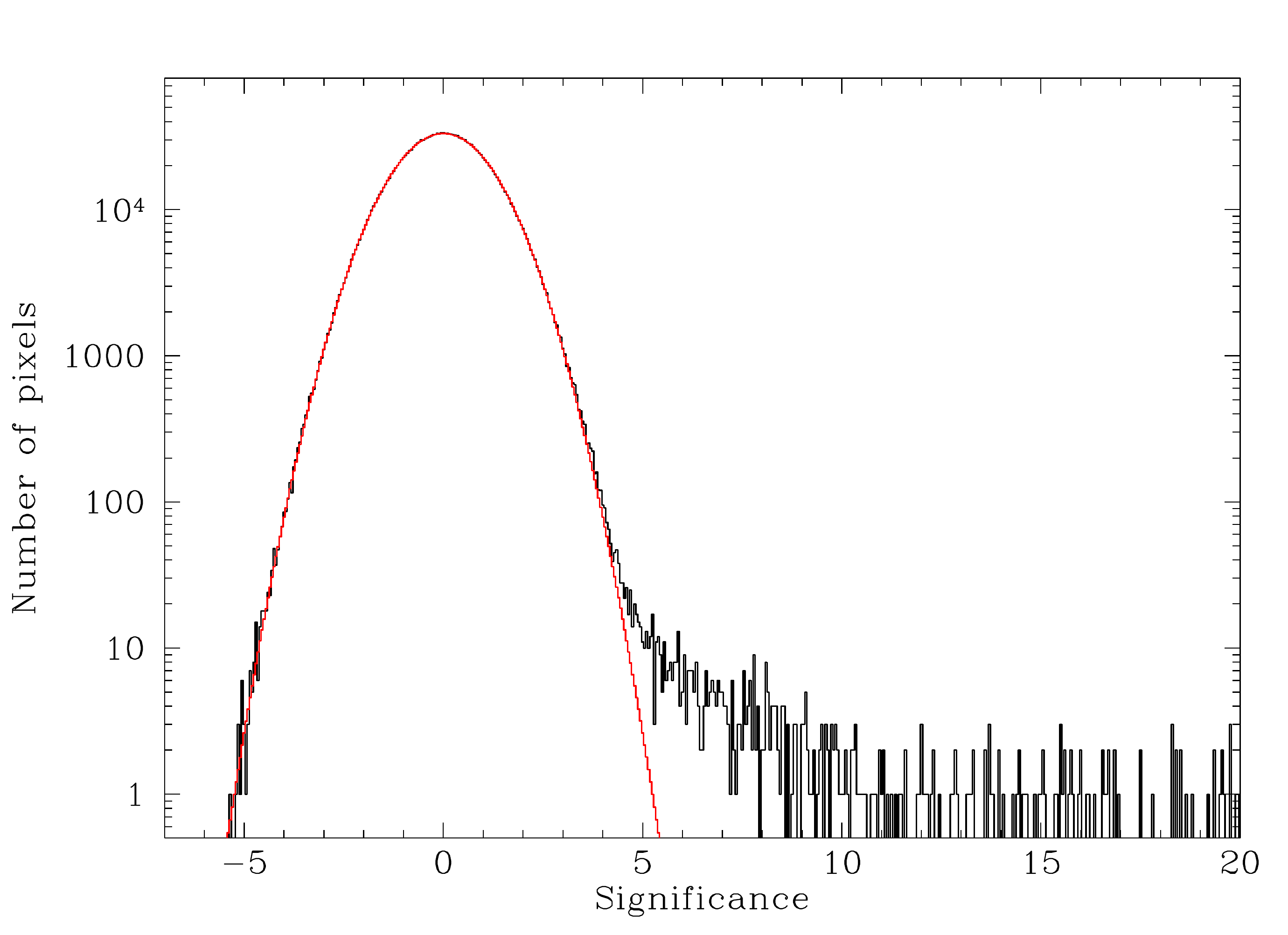}~\includegraphics[width=.48\textwidth]{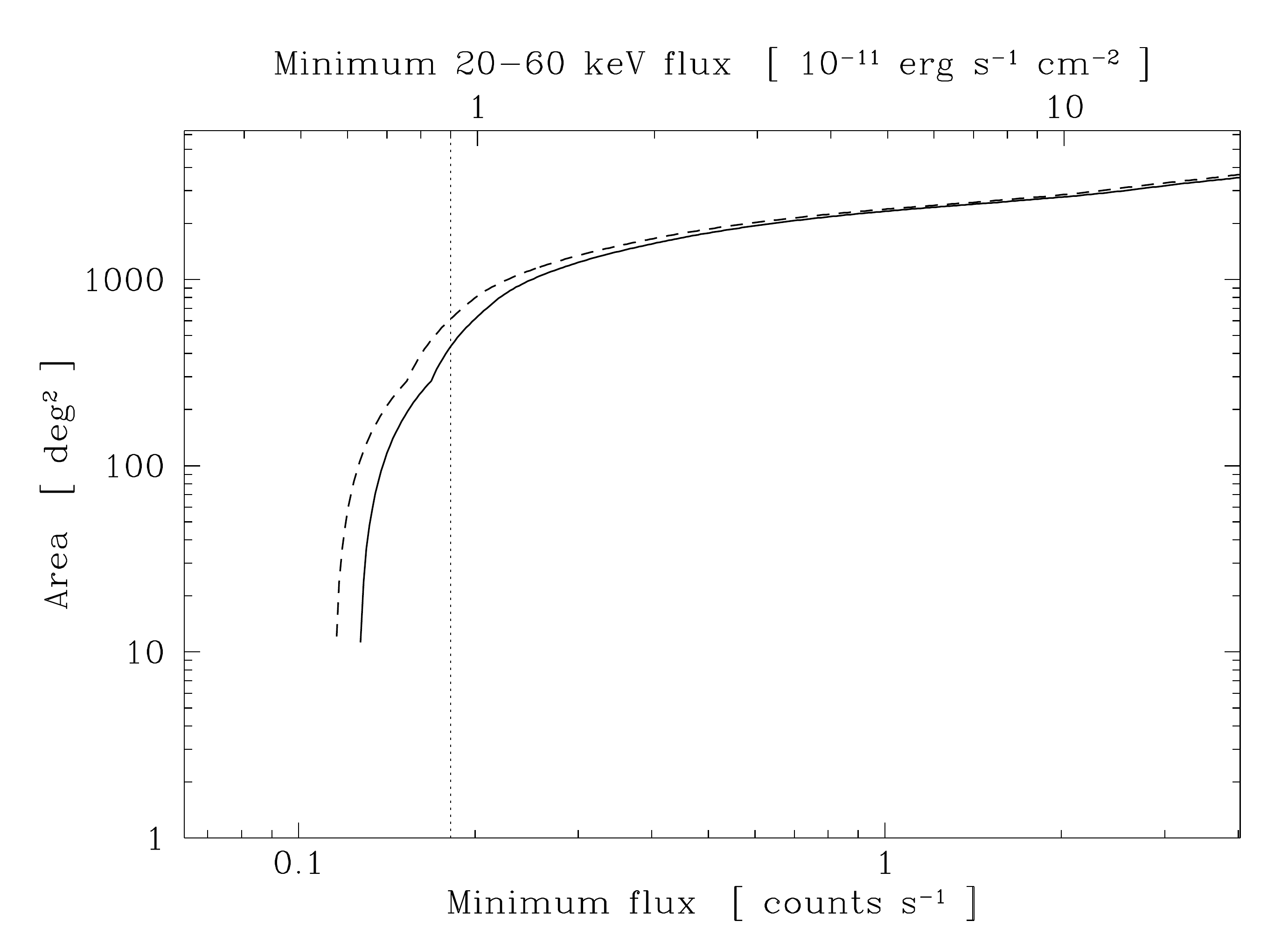}
 \caption{Properties of the mosaic of the 3C~273/Coma region. Left: Histogram of the pixels' significance. The red curve is a Gaussian fit to the part of the histogram with $\sigma\le 3$. Right: Surface of the sky over which a given flux results in a significance $\sigma\ge 5.5$ (solid line) and $\sigma\ge 5.0$ (dashed line). The vertical dotted line shows the sensitivity limit of other hard X-ray surveys.}
 \label{prop}
\end{center}
\end{figure}
Several INTEGRAL core-programme and open-time observations have covered the sky region around Coma cluster and 3C~273, which is illustrated on Fig.~\ref{survey}. We selected all available INTEGRAL pointings within 30 degrees of a position located between these two sources, which resulted in 1660 pointings for a total elapsed observing time of 3\,936\,234\,s and a dead-time corrected good exposure of 2\,733\,202\,s. Most pointings belong to four 5x5 dithering patterns repeated several times, plus a specific rectangular pattern used during the core-programme observation.

Sky images in the 20-60\,keV energy ranges were created from the data taken by the ISGRI detector of the IBIS imager on board INTEGRAL \cite{UberEtal-2003-IbiIma}. The 3000x3000-pixel resulting mosaic image was built in equatorial coordinates with a tangential projection using a factor-2 over-sampling when compared to the individual input sky images; this results in a pixel size of 2.4 arcmin in the center of the mosaic and of about 1.6 arcmin in the outskirts of the image, roughly 40 degrees away from the center.

To avoid any systematic bias in the pixel significance distribution, we excluded the area for which the effective exposure time turned out to be smaller than 10\,ks. The resulting pixel significance distribution is shown on Fig.~\ref{prop} left. The part of the histogram with significance smaller than 3 is perfectly modeled with a Gaussian distribution. The sigma of the Gaussian is however 1.1, instead of the expected value of 1. The centroid is found at a value of 0.009, which has negligible impact on the analysis.

We extract the fluxes and significance of the candidate sources by fitting a PSF directly over the mosaic. In order to derive the sensitivity as a function of effective exposure time, we extracted fake sources at random over the mosaic and estimated their significance distribution. Taking into account repetitions (our mosaic contains about 13\,800 independent pseudo-sources), we obtain that a source with a $5\sigma$ significance has only 20\% chances of being real, and the figure increases to 85\% if we set the threshold to $5.5\sigma$.

\section{Properties of the hard X-ray AGN population}
We find 34 candidate sources in the mosaic with significance larger than $5\sigma$. Source parameters are listed in \cite{PaltEtal-2008-DeeInt}. Taking into account the probabilities that some sources are fake, we end up with approximately 26 real sources. 22 of these sources are found to coincide with known low-redshift ($z<0.05$) AGN. We found that the probability of such chance coincidence is of the order of 1\%. Thus we can consider that these 22 sources are real. We nevertheless use the formal probabilities of the sources being real in all analyzes of the AGN population below.

\subsection{Hydrogen column density distribution}
\begin{figure}
\begin{center}
 \includegraphics[width=.7\textwidth]{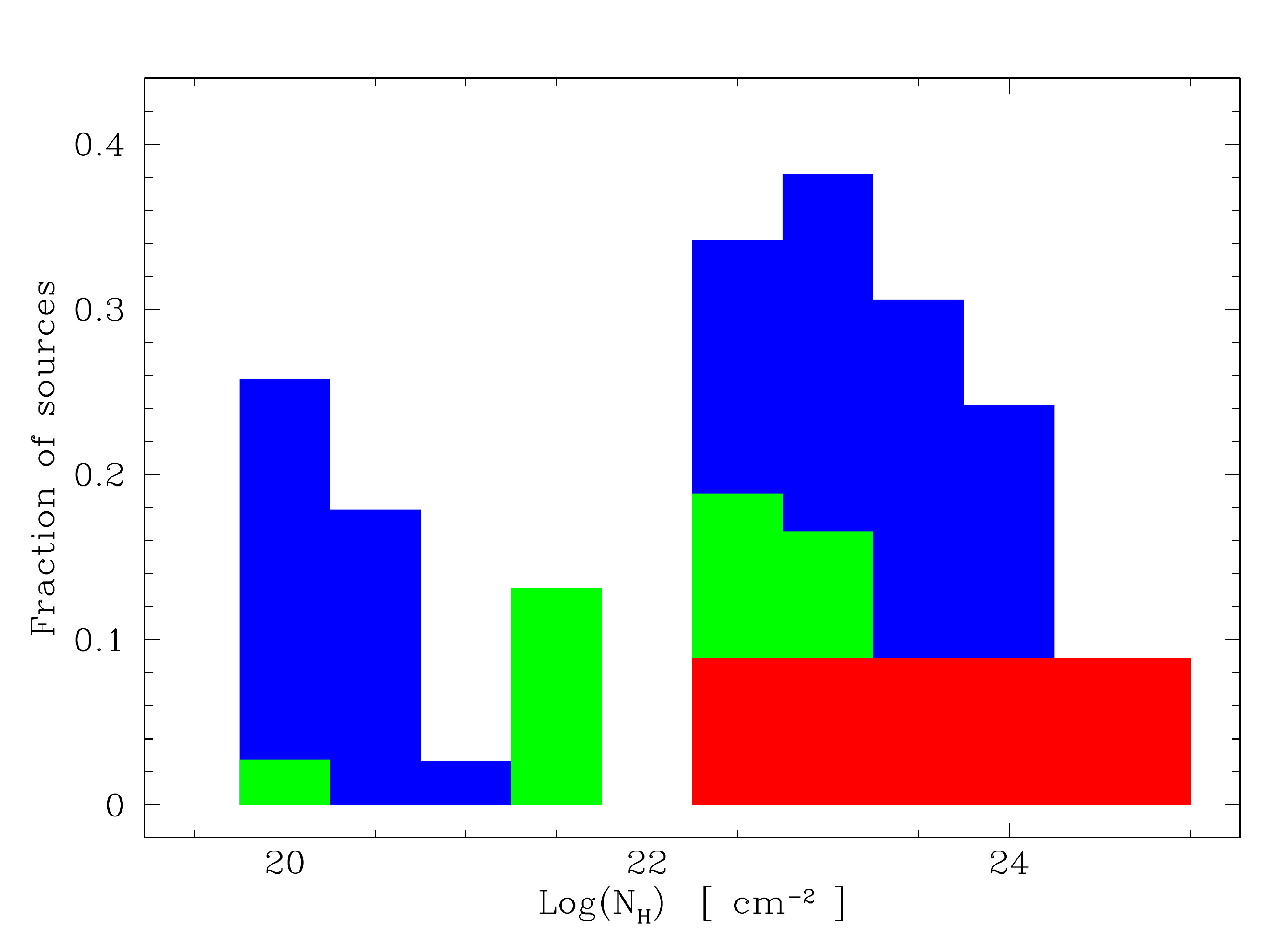}
 \caption{Histogram of the \nh\ distribution for the candidate sources. The blue histogram shows sources with measured intrinsic \nh's. The green histogram shows the sources whose \nh\ is derived from the presence of a counterpart in the RASS-BSC. The red histogram shows the sources without counterparts in the RASS-BSC, with log\,\nh\ evenly distributed between $22.25$ and $25$.}
 \label{nh}
\end{center}
\end{figure}
17 of the 34 candidate sources have adequate X-ray measurements allowing us to determine their hydrogen column density, \nh. For the 17 remaining objects, we check whether there is a counterpart in the ROSAT all-sky survey bright source catalog (RASS-BSC) \cite{VogeEtal-1999-ROSBSC}. This catalog is quasi-complete over the full sky down to a flux of 0.05\,cts\,s$^{-1}$ in the 0.1--2.4\, keV band. We calculate the expected count rate in the ROSAT band for different values of intrinsic \nh\ assuming a power-law intrinsic emission with index $\Gamma=1.9$. While the method is rather crude, it is sufficient to obtain a moderately accurate estimate of \nh. In particular, we checked that it is able to identify correctly, for all 17 sources with \nh\ measurements, whether an AGN is absorbed (\nh$>10^{22}$\,cm$^{-2}$) or not. When the source is not detected in RASS-BSC, \nh\ is roughly constrained within the range $2\times 10^{22}$--$10^{25}$\,cm$^{-2}$.

Figure~\ref{nh} shows the distribution of the intrinsic hydrogen column density \nh\ for the 34 sources detected in the mosaic and weighted by their probabilities. The fraction of absorbed objects, i.e.\ those with \nh$>10^{22}$\,cm$^{-2}$ is found to be 70\%; if one discards the 11 sources without X-ray counterparts, this figure becomes 46\%, making it a stringent lower limit. None of the 23 sources with measured or estimated \nh\ are Compton-thick (\nh$>10^{24}$\,cm$^{-2}$), placing an upper limit to the fraction of Compton-thick objects of 24\%. These figures are consistent with those found in previous hard X-ray surveys. In a recent modeling of the cosmic X-ray background \cite{GillEtal-2007-SynCos}, it was predicted that, at the level of $10^{-11}$\,erg s$^{-1}$ cm$^{-2}$, the fraction of absorbed AGN (with \nh$>10^{22}$\,cm$^{-2}$) should be 65\%, which is perfectly compatible with our measurement. Their expected fraction of Compton-thick AGN (15\%) is also compatible with our upper limit, although the lack of detection of any true Compton-thick object makes our constraint rather weak.

We do not observe any drop in the fraction of absorbed objects at high luminosity, contrarily to several observations in the 2--10\,keV range \cite{UedaEtal-2003-CosEvo,LafrEtal-2005-HarXra,GeorEtal-2006-DeeCha}. While this may due to the small hard X-ray sample sizes in general and to the very small number of high-luminosity objects in particular, we point out that currently no hard X-ray survey has observed this effect with some significance.

\subsection{Comparison with the AGN population observed at 2--10\,keV}
\begin{figure}
\begin{center}
 \includegraphics[width=\textwidth]{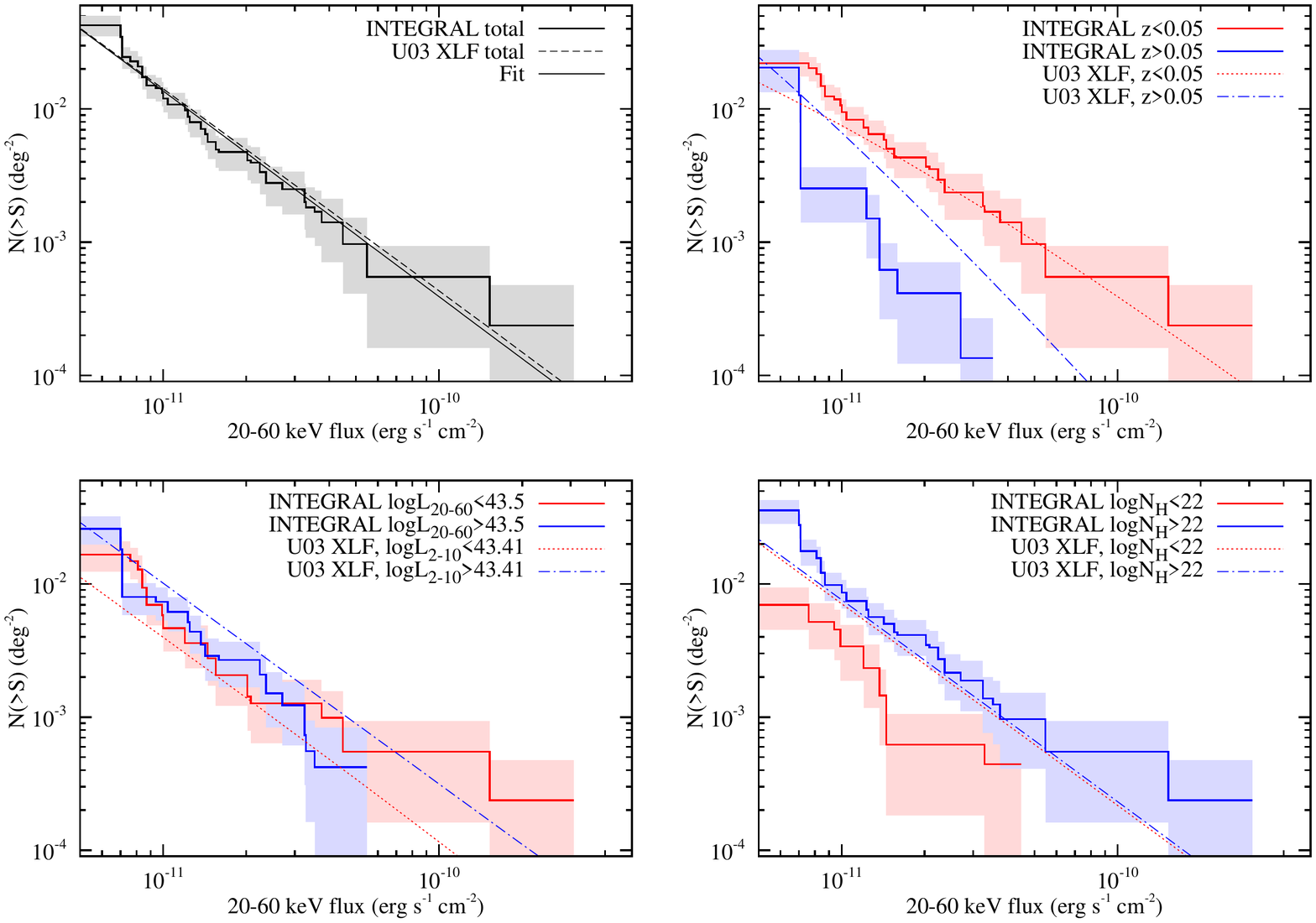}
 \caption{$\log N$--$\log S$ diagrams for the sources detected by INTEGRAL. In all panels the solid lines are from our work and the dashed or dotted line are model extrapolations from \cite{UedaEtal-2003-CosEvo}. Top left: $\log N$--$\log S$ for the full list of candidate sources with the 1$\sigma$ uncertainties. The solid line is the power-law best fit. Top right: $\log N$--$\log S$ in two redshift bins; red: $z < 0.05$; blue: $z>0.05$. Bottom left: $\log N$--$\log S$ in two luminosity bins; red: $L_{20-60\,\mathrm{keV}} < 10^{43.5}$; blue: $L_{20-60\,\mathrm{keV}} > 10^{43.5}$.
Bottom right: $\log N$--$\log S$ in two \nh\ bins; red: \nh$< 10^{22}$\,cm$^{-2}$; blue: \nh$> 10^{22}$\,cm$^{-2}$.`U03' refers to \cite{UedaEtal-2003-CosEvo}.}
 \label{models}
\end{center}
\end{figure}
In Fig.~\ref{models} top left we compare our source counts to the predictions of the 2--10\,keV population model of \cite{UedaEtal-2003-CosEvo}.  We predict the 20--60\,keV source counts by integrating the 2--10\,keV model AGN population over the $0 < z < 2$, $10^{41} < L_{2-10\,\mathrm{keV}} < 10^{47}$\,erg s$^{-1}$ range. The conversion from rest-frame intrinsic (i.e. before absorption) 2--10\,keV luminosity to observed frame 20--60\,keV flux is made using the same spectral model as in \cite{UedaEtal-2003-CosEvo}, namely a power-law spectrum with $\Gamma = 1.9$, a cut-off rest-frame energy of 500\,keV, and a reflection component from cold material. With this spectral model, $S_{20-60\,\mathrm{keV}} = 1.24 \times L_{2-10\,\mathrm{keV}}/4 \pi d^{2}_{\mathrm L}$ at redshift $\sim 0$ ($d_{\mathrm L}$ is the luminosity distance in cm). We use here the model of \cite{UedaEtal-2003-CosEvo} which includes a mix of unabsorbed and Compton-thin sources, but no Compton-thick sources; the Compton-thick population has indeed not been measured and has been treated somewhat arbitrarily by adding a number of these objects equivalent to that of the Compton-thin ones. We have therefore assumed that absorption effects on this population in the 20--60\,keV band are negligible.

Under these assumptions, we can see that the 2--10\,keV model provides a good match to both the slope and normalization of the total 20--60\,keV source counts (see Fig.~\ref{models} top left),
leaving little room for a significant additional population of moderately Compton-thick sources. We examine the population in more detail by splitting the sample into low ($z<0.05$) and high redshift sources, low ($L_{20-60\,\mathrm{keV}} < 10^{43.5}$\,erg s$^{-1}$) and high luminosity sources, and low (\nh$< 10^{22}$\,cm$^{-2}$) and high absorption sources. In Fig.~\ref{models} top right we show the source counts and model predictions for redshifts either below, or above 0.05. Redshifts are unknown for 6 sources. The observed counts and model predictions agree reasonably well given the relatively small numbers of observed sources. Above $S_{20-60\,\mathrm{keV}} \sim 10^{-11}$\,erg cm$^{-2}$ s$^{-1}$, the 2--10\,keV model from \cite{UedaEtal-2003-CosEvo} overpredicts the number of observed sources with $z>0.05$.

In Fig.~\ref{models} bottom left we show the source counts and model predictions for observed sources above and below a luminosity of $L_{20-60\,\mathrm{keV}} = 10^{43.5}$\,erg s$^{-1}$. For our given spectral model this corresponds to $L_{2-10\,\mathrm{keV}} = 10^{43.41}$\,erg s$^{-1}$ (as described above). The observed counts and model predictions are roughly in agreement, given the small number of sources.

Absorbing column estimates or lower limits are available for our entire sample and so, in Fig.~\ref{models} bottom right, we show the source counts and model predictions separated into sources with \nh\ either greater than, or less than $10^{22}$\,cm$^2$. Here it is clear that the 2--10\,keV model is a poor predictor of the observed source counts. We see that the more absorbed sources constitute at least 2/3 of the total 20--60\,keV source counts over the flux range of the sample, whereas the 2--10\,keV model of \cite{UedaEtal-2003-CosEvo} predicts equal number of absorbed and non-absorbed sources over the luminosity and redshift range probed by the INTEGRAL observations. Taken at face value, it means that the observed and model $N_H$ distributions differ significantly. A follow-up with more sensitive medium energy X-ray observations is therefore needed to confirm this result, especially for the sources with no direct \nh\ measurement.

\section{The hard X-ray luminosity function}
\begin{figure}
\begin{center}
 \includegraphics[width=.7\textwidth]{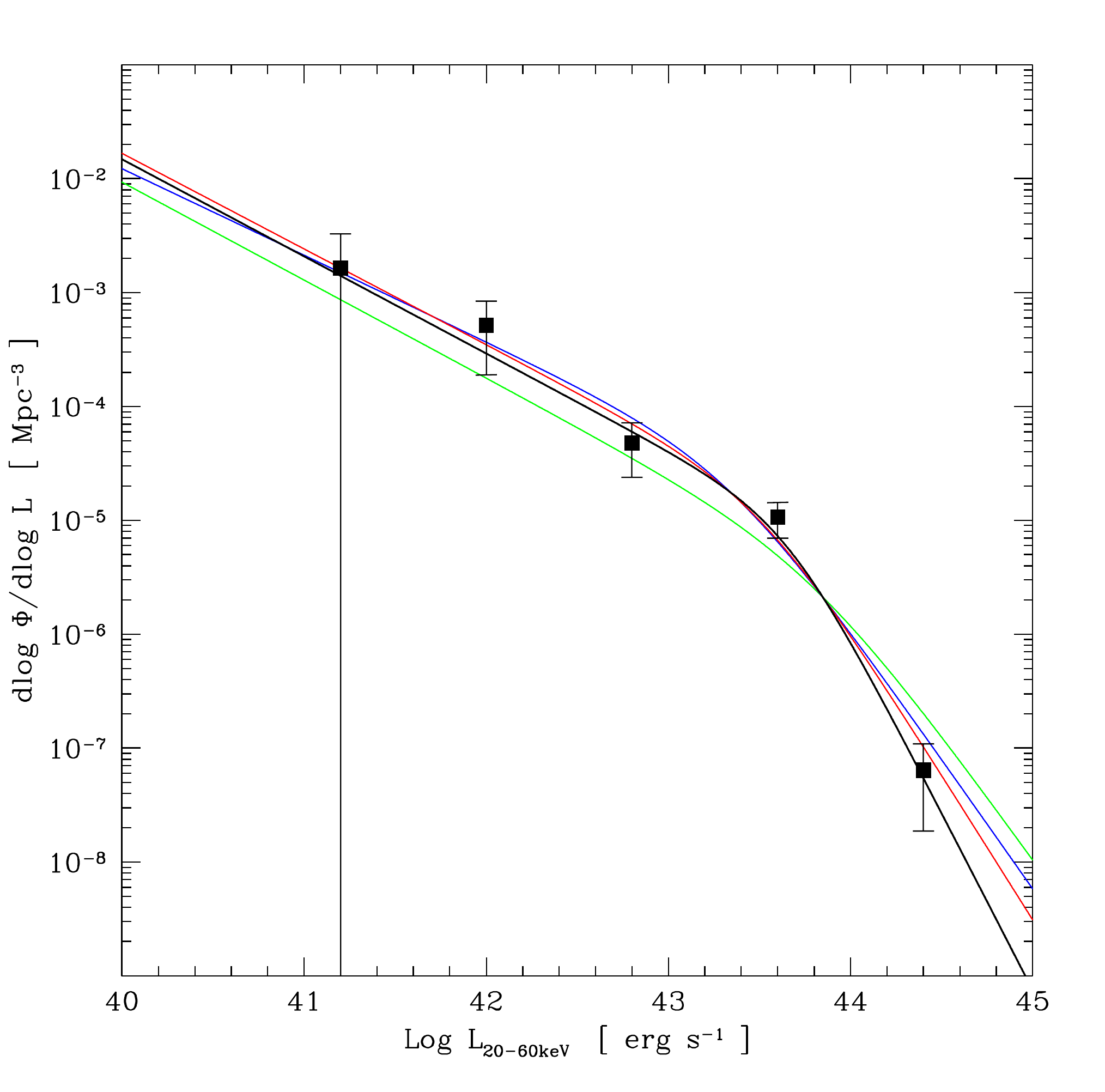}
 \caption{Hard AGN X-ray luminosity function from the 3C\,273/Coma survey. The black points use the $V/V_{\mathrm{max}}$ estimator, while the black curve has been obtained with a maximum likelihood. The blue curve is the LF from an INTEGRAL all-sky survey \cite{SazoEtal-2007-HarXra} and the red curve is from the Swift/BAT survey \cite{TuelEtal-2008-SwiBat}. The LF from the luminosity-dependent density evolution model found in the X-rays \cite{UedaEtal-2003-CosEvo} and converted to the hard X-ray domain using $\Gamma=1.9$ is shown in green.}
 \label{fig0}
\end{center}
\end{figure}
We use the candidate sources firmly detected as AGN to determine the AGN luminosity function (LF) in the 20--60\,keV energy range. We use only objects with redshifts $0\le z\le 0.05$ in order to study the local AGN population with negligible evolution. Because of the small number of sources, we use a parametric method to derive the LF. We assume the standard AGN luminosity function:
\begin{equation}
\frac{{\mathrm d}\Phi(L)}{{\mathrm d}\log L}=\frac{\Phi^*}{(L/L^*)^{\alpha}+(L/L^*)^{\beta}},
\label{eq:lf}
\end{equation}
which describes a broken power-law, changing from index $\alpha$ to index $\beta$ at characteristic luminosity $L^*$. The parameters $L^*$, $\alpha$ and $\beta$ parameters are determined using a maximum-likelihood (ML) test, based on the idea proposed by \cite{SandEtal-1979-VelFie}, modified to take into account our sensitivity map. ML-based methods are insensitive to sample sizes, therefore $\Phi^*$ cannot be estimated this way. Thus we also estimate the LF using the standard non-parametric $V/V_{\mathrm{max}}$ estimator. $\Phi^*$ is then obtained by fitting the $V/V_{\mathrm{max}}$ LF estimates with the $\Phi(L)$ distribution from Eq.~(\ref{eq:lf}), letting only $\Phi^*$ free and fixing the other parameters to the ML values. We obtain the following results for the four LF parameters:
\begin{equation}
 \begin{array}{ccl}
  \log L^*&=&43.66\,^{+0.28}_{-0.60}{\mathrm {~erg~s}}^{-1}\\[1mm]
  \alpha&=&\z 0.85\,^{+0.26}_{-0.38}\\[1mm]
  \beta&=&\z 3.12\,^{+1.47}_{-1.02}\\[1mm]
  \Phi^*&=&\z 1.12\,^{+5.04}_{-0.77}~10^{-5}{\mathrm {~Mpc}}^{-3}\\
\end{array}
\end{equation}
As can be seen from Fig.~\ref{fig0}, the AGN hard X-ray luminosity functions from INTEGRAL all-sky survey \cite{SazoEtal-2007-HarXra} and the Swift/BAT survey \cite{TuelEtal-2008-SwiBat} are perfectly compatible with our estimate.

$\Phi^*$ and $L^*$ are very strongly correlated; therefore the LF normalization is known with a much better accuracy than the uncertainty on $\Phi^*$ suggests. As an example, the luminosity density integrated above $L_{20-60\,\mathrm{keV}}=10^{41}$\,erg s$^{-1}$ is:
\begin{equation}
 \begin{array}{ccl}
  W(L>10^{41})&=&\int_{41}^\infty L\frac{\mathrm{d}\phi}{\mathrm{d}\log L}\,\mathrm{d}\log L =\\[2mm]
  &=&\z0.90\,^{+0.19}_{-0.25}~10^{39}{\mathrm {erg~s}}^{-1}\mathrm{{~Mpc}}^{-3}
\end{array}
\end{equation}
The luminosity density is again perfectly compatible with \cite{SazoEtal-2007-HarXra} and \cite{TuelEtal-2008-SwiBat} ($W(L>10^{41})=1.02~10^{39}$ and $1.03~10^{39}$ erg s$^{-1}$ Mpc$^{-3}$ respectively).

Figure~\ref{fig0} left also shows the local, $z=0$, X-ray LF from \cite{UedaEtal-2003-CosEvo}. This LF is significantly below the hard X-ray ones by a factor 1.5. There are two extreme ways to reconcile the LFs: either the X-ray LF is shifted vertically, which correspond to a change in AGN density, or horizontally, which corresponds to a change in AGN luminosity.

A change in density can be explained if there is a fraction of objects that have been completely missed in X-ray surveys, for instance the Compton-thick population. This would imply that one third of the AGN in our sample are Compton-thick, which is above our upper limit.

Alternatively, a change in luminosity resulting in a horizontal shift in the LF might be explained if we underestimated the hard X-ray luminosity of AGN observed in the X-rays. This is perfectly possible, since we had to transform the unabsorbed 2--10\,keV X-ray luminosity assuming a spectral shape, which we chose as a simple power-law with photon index $\Gamma=1.9$. The LFs would become consistent if we choose instead a photon index $\Gamma=1.7$. Interestingly, we obtain a very good approximation to a power-law with photon index $\Gamma=1.7$ if we combine a power-law with photon index $\Gamma=1.9$ with a reflection with reflection fraction $R=1$. Thus the comparison of the LFs is evidence that large reflection fractions with $R\sim 1$ are common in AGN.

\section{Prospects for an ultra-deep INTEGRAL extragalactic survey}
\begin{figure}
\begin{center}
 \includegraphics[width=.7\textwidth]{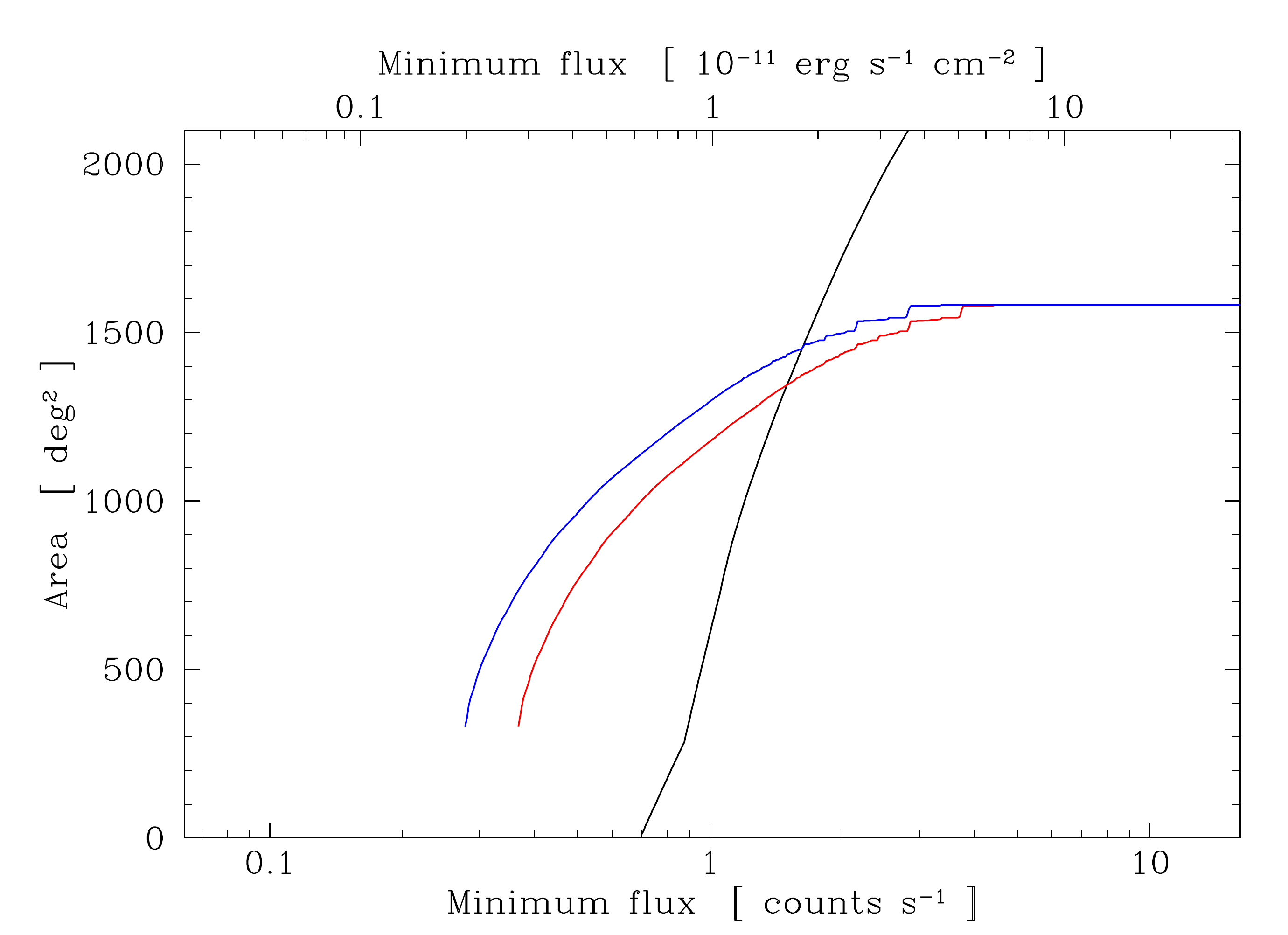}
 \caption{Surface of the sky over which a given flux results in a significant detection. The red curve corresponds to a 5x5 dithering pattern with a 10\,Ms total exposure time. The blue curve corresponds to the same observing pattern, but with a 20\,Ms exposure time. The black curve is the 3C\,273/Coma survey \cite{PaltEtal-2008-DeeInt}}
 \label{fig1}
\end{center}
\end{figure}
\begin{figure}
\begin{center}
 \includegraphics[width=.48\textwidth]{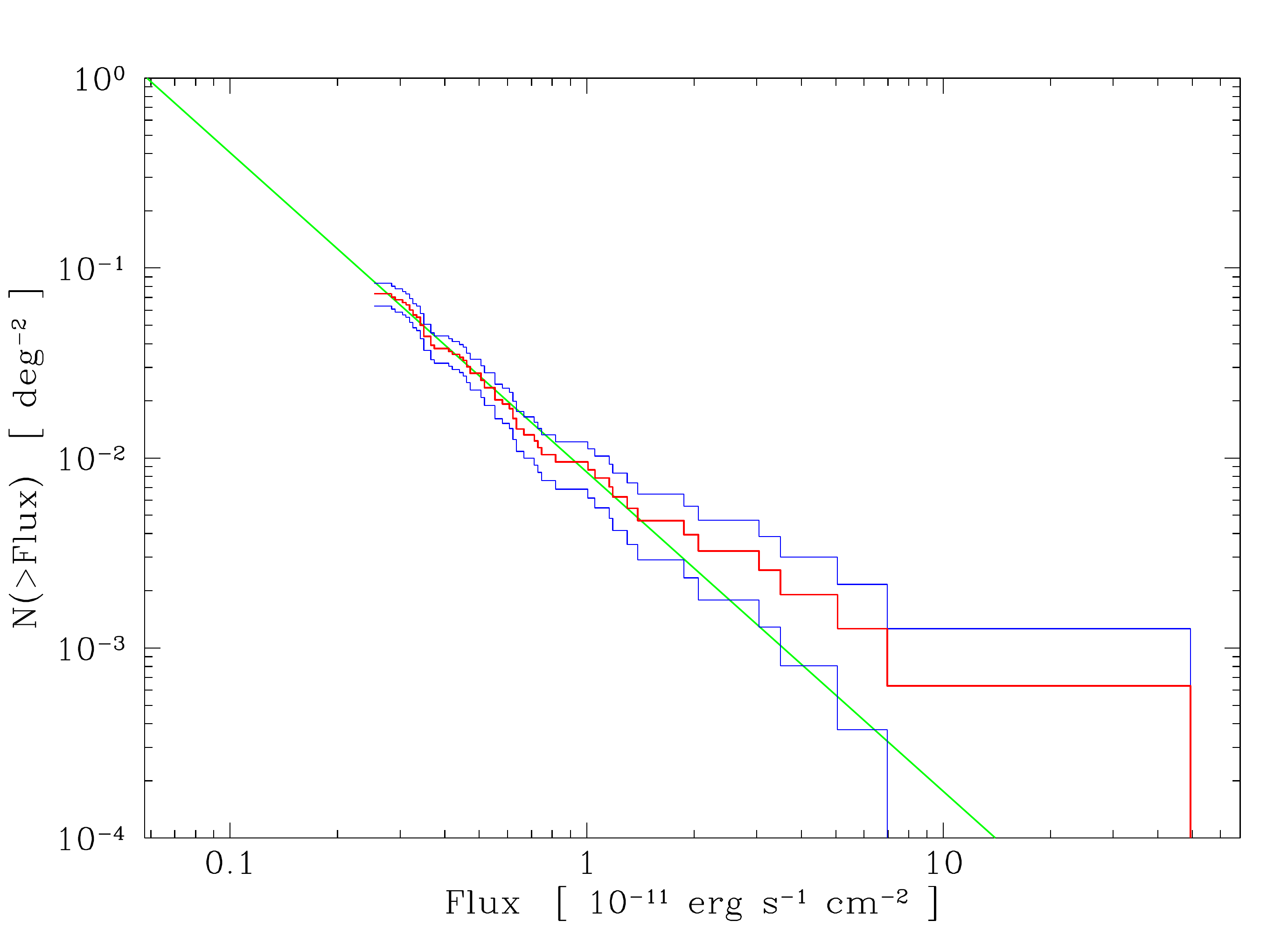}~\includegraphics[width=.48\textwidth]{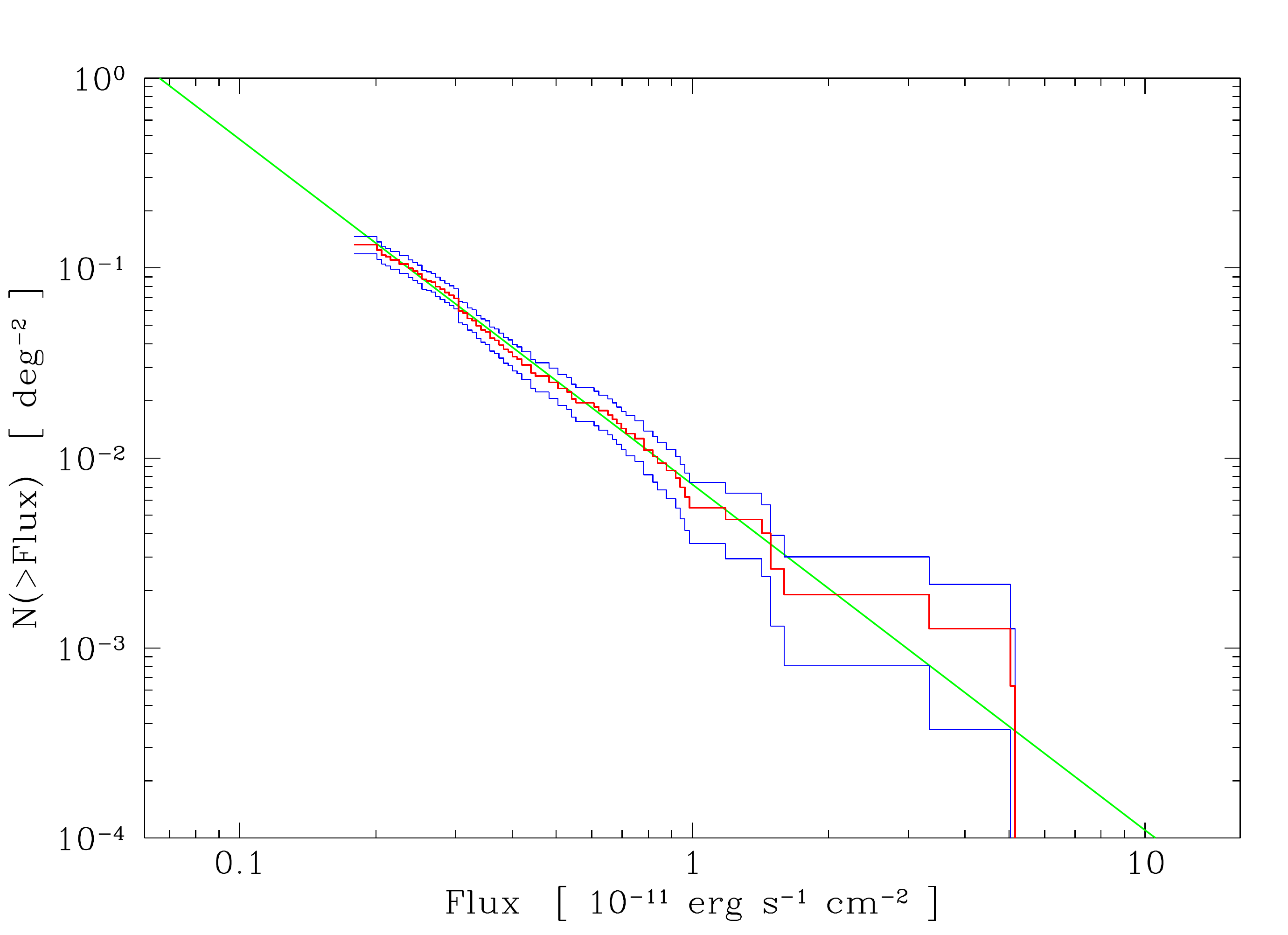}
 \caption{Simulated Log\,$N$--Log\,$S$ diagram (red) resulting from a 5x5 dithering pattern with a 10\,Ms total exposure time (left) and a 20\,Ms total exposure time (left) respectively. The blue line show the 1$\sigma$ uncertainties. The green line is a power-law fit to the Log\,$N$--Log\,$S$ diagram.}
 \label{fig2}
\end{center}
\end{figure}

In terms of sensitivity, coded-mask telescope performances are well below those of focusing telescopes. While telescopes able to focus hard X-rays will fly in a few years and will make wonderful contributions to the study of the AGN population, their fields of view are much smaller than those of Swift/BAT or INTEGRAL/IBIS, which limits their survey capabilities. It is therefore worth exploring the prospects of an ultra-deep INTEGRAL/IBIS extragalactic observation that could realistically be completed before the end of the mission.

We simulated INTEGRAL/IBIS observations following a 5x5 dithering pattern with 10\,Ms and 20\,Ms total exposure times. Fig.~\ref{fig1} shows the surface of the sky over which a given flux sensitivity would be reached with such observations. On the basis of these curves, we can draw a population of AGN assuming that the Log\,$N$--Log\,$S$ diagram obtained in the 3C\,273/Coma survey. Log\,$N$--Log\,$S$ diagrams of such simulated populations are shown on Fig.~\ref{fig2}.

The average number of sources is 57 for the 10\,Ms observation and 95 for the 20\,Ms one, with 0-area minimum flux limits of 3 and $2\times 10^{-12}$\,erg s$^{-1}$ cm$^{-2}$ respectively between 20 and 60\,keV. Below $10^{-11}$ erg s$^{-1}$ cm$^{-2}$, AGN with redshifts $z>0.05$ dominate the full population. We find that at least 20 objects with $z>0.05$ will be found in the 10\,Ms observation, and at least 50 in the 20\,Ms one.

We can translate these sample sizes in uncertainties on the normalization of the luminosity function. We obtain an uncertainty better than 22\% with the 10\,Ms observation and better than 14\% with the 20\,Ms one. The X-ray luminosity function normalization increases very rapidly, like $(1+z)^4$, at very small redshifts \cite{UedaEtal-2003-CosEvo}, implying a 20\% density evolution between $z=0.025$ and $z=0.075$. As the local hard X-ray luminosity function will be quite well constrained at the end of the Swift mission, this means that an ultra deep INTEGRAL/IBIS extragalactic observation has the potential to constrain AGN density evolution in the hard X-rays.

\section{Conclusion}
While the extragalactic hard X-ray survey presented here covers an area much smaller than other recent similar surveys \cite{SazoEtal-2007-HarXra,TuelEtal-2008-SwiBat}, we reached limiting fluxes a factor 1.5 times deeper. We obtained a source density of 0.013\,deg$^{-1}$ at a 20--60\,keV flux level of $10^{-11}$\,erg cm$^{-2}$ s$^{-1}$. We resolved approximately 2.5\% of the cosmic X-ray background. Comparison with the 2--10\,keV domain shows that the total Log\,$N$--Log\,$S$ diagrams are compatible, indicating that there isn't any large population of bright Compton-thick objects missed in the 2--10\,keV surveys and appearing in the hard X-rays.

No object is our sample is truly Compton-thick, although the upper limit to the Compton-thick fraction is compatible with the fraction expected from models of the cosmic X-ray background \cite{GillEtal-2007-SynCos}. Nevertheless the absence of such objects makes that the case for the existence of such population at the required level is rather weak. Follow-up observations of our sources without adequate \nh\ measurements are in progress, and may solve the puzzle. It seems however quite probable that we shall end-up with conflicting \nh\ distributions, which will have to be explained.

We present a truly local hard X-ray luminosity function ($z<0.05$). We find a LF quite compatible with the latest all-sky INTEGRAL and SWIFT surveys. A discrepancy is however found between the hard X-ray LFs and the 2--10\,keV ones, which we interpret as an excess luminosity in the hard X-rays, which could result from a reflection hump with reflection fraction $R\sim 1$. 

While still suffering from low sensitivities compared to their modern counterparts working in the soft and medium X-rays, INTEGRAL and Swift are the only instruments currently available to perform surveys above 15\,keV. The importance of this energy domain, which is unaffected by obscuration below $\sim 10^{25}$\,cm$^{-2}$, is such that efforts to build statistically representative samples must be pursued. While the Swift mission will cover the full sky with quite deep observations, it is possible for INTEGRAL to perform ultra-deep observations way beyond what could be obtained with Swift. We showed here that such INTEGRAL observations would produce numerous faint sources allowing us to quantify the AGN population with very good accuracy. In addition they would allow us to explore the evolution of the AGN population in the hard X-rays for the first time. Therefore ultra-deep INTEGRAL high-latitude observations, maybe reaching some 20\,Ms total exposure time, have a very strong scientific potential and should absolutely be performed before the end of the INTEGRAL mission.

\bibliographystyle{JHEP}
\bibliography{biblio}

%

\end{document}